\documentstyle[preprint,aps]{revtex}
\def\be{\begin{equation}}
\def\ee{\end{equation}}
\def\bea{\begin{eqnarray}}
\def\eea{\end{eqnarray}}
\def\bem{\begin{mathletters}}
\def\eem{\end{mathletters}}
\def\A{{\cal A}}
\def\l{\lambda}
\def\M{\bbox{\cal M}}
\def\m{\mu}
\def\e{\epsilon}
\def\d{\delta}
\def\n{\nu}
\def\k{\kappa}
\def\G{\Gamma}
\def\half{\frac{1}{2}}
\def\a{\alpha}
\def\L{{\cal L}}
\def\H{\bbox{\cal H}}
\def\t{\tau}
\def\S{\bbox{\Psi}}

\begin{document}
\draft
\title{ABOUT SOME PROBLEMS RAISED BY THE RELATIVISTIC FORM OF DE-BROGLIE--BOHM THEORY OF PILOT WAVE}
\author{Ali Shojai\thanks{Email: SHOJAI@THEORY.IPM.AC.IR}}
\address{Physics Department, Tarbiat Modares University, P.O.Box 14155--4838, Tehran, IRAN.}
\address{\&}
\address{Institute for Studies in Theoretical Physics and Mathematics, P.O.Box 19395--5531, Tehran,
IRAN.}
\author{Fatimah Shojai\thanks{Email: FATIMAH@THEORY.IPM.AC.IR}}
\address{Physics Department, Iran University of Science and Technology, P.O.Box 16765--163, Narmak, Tehran, IRAN.}
\address{\&}
\address{Institute for Studies in Theoretical Physics and Mathematics, P.O.Box 19395--5531, Tehran,
IRAN.}
\maketitle
\begin{abstract}
The standard relativistic de-Broglie--Bohm theory has the problems of tacyonic solutions and the incorrect non--relativistic limit. 
In this paper we obtain a relativistic theory, not decomposing
the relativistic wave equations but looking for a generalization of non--relativistic Bohmian
theory in such a way that the correct non--relativistic limit emerges. In this way we are able to construct a relativistic de-Broglie--Bohm theory both for a single particle and for a many--particle
system. At the end, the theory is extended to the curved space--time and the connection with
quantum gravity is discussed.
\end{abstract}
\section{Introduction}
Bohmian mechanics\cite{Bohm1,Bohm2,Bohm3,Holland}
 is an attempt  to make a causal deterministic theory in such a way that all
experimentally approved results of quantum mechanics would be reproduced. The main character
of Bohm's theory is the {\it quantum potential\/}. In this theory it is assumed that in addition to 
the classical potential there is another potential, quantum potential, which can be derived from the 
wave function given by:
\be
Q=-\frac{\hbar^2}{2m}\frac{\nabla^2|\S|}{|\S|}
\label{QP}
\ee
Respectively, there is a quantum force derivable from the quantum potential as:
\be
\vec{F}_Q=-\vec{\nabla}Q
\ee
The $\S$--field itself satisfies the Schr\"odinger equation:
\be
i\hbar\frac{\partial\S}{\partial t}=-\frac{\hbar^2}{2m}\nabla^2\S+V\S
\ee
All the magic behaviour of quantum particles can be described in terms of this force.
The theory can simply extended to the case of many--particle systems. The quantum potential
is given by:
\be
Q=-\frac{\hbar^2}{2}\sum_{i=1}^N\frac{\nabla_i^2|\S|}{m_i|\S|}
\ee
while $\S$ satisfies the many--particle Schr\"odinger equation.

In fact the above interpretation rests on the fact that when one decomposes the $\S$--field
in the Schr\"odinger equation into its norm and phase, one gets two equations: the first is the 
Hamilton--Jacobi equation including the quantum potential in which $\hbar$ times the phase
of the wavefunction is the Hamilton--Jacobi function, and the second is the continuity equation
if one assumes that the square of the norm of the wavefunction is equal to the density of 
an hypothetical ensemble of the particle under consideration.

Although Bohmian mechanics as stated above is non-relativistic, about one quarter of a century before Bohm, de-Broglie presented a theory where Bohm's theory is a special limit of it\cite{de1,de2,de3}. The important point is that that theory was written in a relativistic form.

In fact if one takes the Klein--Gordon equation and then decomposes the wavefunction 
into its phase and norm, one gets the relativistic Hamilton--Jacobi and continuity equations:
\be
\partial_\m S\partial^\m S=m^2c^2+\hbar^2\frac{\Box|\S|}{|\S|}=\M^2c^2
\ee
\be
\partial_\m(|\S|^2\partial^\m S)=0
\ee
in which $S$, the Hamilton--Jacobi function, is in fact $\hbar$ times the phase of the wavefunction.
$\M$ can be interpretated as the {\it quantum mass\/}, which is in fact a field. de-Broglie theory has problems.\cite{Holland}
 These problems 
are in fact related to the problems of the standard relativistic quantum mechanics.
Difficulties are related to the definition of $\M$. 
In general, $\M^2$ is not positive--definite.\cite{Holland} Accordingly this theory admits tachyonic solutions which may be argued that this is a bad property.

In fact we have two ways for understanding Bohm's theory. First, one can say that the wavefunction 
is the most important quantity, and the Bohmian trajectories can be obtained by decomposing the
wavefunction into its norm and phase, and interpretating them as above. The second way is to 
think about the wavefunction as an objectively real field exerting the quantum force on the 
particle. In the first approach we have the Schr\"odinger equation (in the non--relativistic regime), as
the essential equation of the theory plus the decomposition of the wavefunction as a rule. In the 
second  approach, the Schr\"odinger and the Newtons second law (containing the quantum force)
are the equations of motion of particle and field. There is no difference between these two approaches
in the non--relativistic domain, so this point is not highlighted in the literature. But for relativistic
systems some differences would emerge.  In the first approach one has the Klein--Gordon equation,
and decomposing it we shall get the above equations of relativistic theory, which has the 
problem of non--positive--definiteness. In the second approach we would accept the Klein--Gordon 
equation as  the field equation, and must write down an equation of motion for the particle which is
maniffestly covariant and which has the correct nonrelativistic limit. We shall show in this paper 
that the result would be different from the first approach, and that there is no non--positive--definiteness problem.

In this paper we shall assume that it is not the correct way only to decompose the wavefunction 
into its phase and norm and then interpret the result. But, in fact, the only lesson from the
non--relativistic theory is that {\it there is some quantum potential with the non--relativistic
limit of the form of equation (\ref{QP})\/}. So one must write down some relativistic equations 
with the correct non--relativistic limit (the non--relativistic Bohmian equation).

Now we have a critreria for writting de-Broglie--Bohm theory in the relativistic domain. It is: {\it A good
relativistic causal quantum mechanics is that one which has the correct nonrelativistic limit and which
is manifestely covariant\/}.
\section{Relativistic Equations of Motion of a Quantum Particle}
The action functional for a single relativistic particle of mass $\M$ (which is a Lorentz invariant
quantity)  is given by:
\be
\A=\int\! d\l \ \half\M(r)\frac{dr_\m}{d\l}\frac{dr^\m}{d\l}
\label{RA}
\ee
in which $\l$ is any scalar parameter used to parametrize the path $r_\m(\l)$. It can be in fact the
proper time $\t$, but in general is different from it.

The equation of motion can be derived from the least action principle. Varying the path:
\be
r_\m\longrightarrow r'_\m=r_\m+\e_\m
\ee
one gets:
\be
\A\longrightarrow\A'=\A+\d\A=\A+\int\! d\l \ \left [ \M\frac{dr_\m}{d\l}\frac{d\e^\m}{d\l}+\half
\frac{dr_\m}{d\l}\frac{dr^\m}{d\l}\e_\n\partial^\n\M\right ]
\ee
According to the least action principle, the correct path satisfies:
\be
\d\A=0\ \ \ \   \text{with fixed boundaries}
\ee
so the equation of motion is:
\be
\frac{d}{d\l}\left (\M u_\m\right )=\half u_\n u^\n \partial_\m\M
\ee
or:
\be
\M\frac{du_\m}{d\l}=\left ( \half\eta_{\m\n}u_\a u^\a-u_\m u_\n\right )\partial^\n\M
\ee
where:
\be
u_\m=\frac{dr_\m}{d\l}
\ee

Let us now investigate the symmetries of the action functional (\ref{RA}).  Suppose we make the following 
transformation:
\be
\l\longrightarrow\l+\d
\ee
In the case of $\l$--translation we have the hamiltonian as the conserved current:
\be
\H=-\L+u_\m\frac{\partial\L}{\partial u_\m}=\half \M u_\m u^\m=E
\ee
This can be seen by setting $\d\A=0$ for $\l$--translations:
\be
0=\d\A=\A'-\A=\int d\l \left ( \half u_\m u^\m u^\n \partial_\n\M+\M u_\m\frac{du^\m}{d\l}\right )\d
\ee
so that the integrand is zero:
\be
\frac{d}{d\l}\left (\half \M u_\m u^\m\right )=0
\ee
This leads to (remembering that the proper time is defined as $c^2d\t^2=dr_\m dr^\m$):
\be
\frac{d\t}{d\l}=\sqrt{\frac{2E}{\M c^2}}
\ee
The equation of motion now can be written as:
\be
\M\frac{dv_\m}{d\t}=\half\left ( c^2\eta_{\m\n}-v_\m v_\n\right )\partial^\n\M
\label{REM}
\ee
where:
\be
v_\m=\frac{dr_\m}{d\t}
\ee

The non--relativistic limit can be derived by letting the particle's velocity be ignorable with repect
to the light's velocity. In this limit the proper time is identical to the time coordinate, $\t=t$. 
The result is:
\be
\m=0\ \ \text{component is satisfied identically.}
\ee
\be
\M\frac{d^2\vec{r}}{dt^2}=-\half c^2\vec{\nabla}\M
\ee
writing the above equation as:
\be
m\frac{d^2\vec{r}}{dt^2}=-\vec{\nabla}\left ( \frac{mc^2}{2}\ln\frac{\M}{\m}\right )
\ee
where $\m$ is an arbitrary mass scale. In order to have the correct limit, the term in the parenthesis 
on the right hand side of the above equation should be equal to the quantum potential given by
the equation (\ref{QP}). So:
\be
\M=\m\exp\left\{-\frac{\hbar^2}{m^2c^2}\frac{\nabla^2|\S|}{|\S|}\right\}
\ee
So the relativistic quantum mass field which is manifestly invariant is given by:
\be
\M=\m\exp\left\{\frac{\hbar^2}{m^2c^2}\frac{\Box|\S|}{|\S|}\right\}
\ee
The first two terms of the above relation are similar to those of the standard relativistic de-Broglie--Bohm 
theory, provided that one sets $\m=m$. So
\be
\M=m\exp\left\{\frac{\hbar^2}{m^2c^2}\frac{\Box|\S|}{|\S|}\right\}
\label{zaq}
\ee
It must be noted here that since the relativistic equation of motion (\ref{REM}) has the correct limit,
the standard relativistic de-Broglie--Bohm theory has not the correct limit as can be checked directly.
In fact if one starts with the standard relativistic theory, and go to the non--relativistic 
limit, one did not get the correct non--relativistic equations.\cite{Thesis}
 This is an additional problem with the
standard relativistic de-Broglie--Bohm theory. In our above formalism this is solved as well as  the
standard problems. Another important point is that the relation (\ref{zaq}) for quantum mass leads to a positive--definite mass square.
\section{Relativistic Equations of Motion of a System of Quantum Particles}
Extension of the results of the previous section to the case of many particle systems is not
very complicated. Simply one can generalize the action functional of a single particle to the 
following action for many particle systems:
\be
\A=\sum_{i=1}^N\int\! d\l_i \ \half\M(r^i_\m)\frac{dr^i_\m}{d\l_i}\frac{dr^{i\m}}{d\l_i}
\ee
in which $\l_i$ represents the parametrization parameter for the $i$th particle, $N$ is number of
particles, and now the 
quantum mass is a function of the location of all the particles. Note that we have considered that 
the mass of each particle is derived from a single field.
The relation of the mass of each particle with this mass field would be derived later.
In order to derive the equations of motion it is more convenient to introduce a new parameter $\l$
and see the $\l_i$'s as functions of this parameter. So the action functional would be:
\be
\A=\int\! d\l \sum_{i=1}^N  \ \half \frac{d\l_i}{d\l}\M(r^i_\m)\frac{dr^i_\m}{d\l_i}\frac{dr^{i\m}}{d\l_i}
\ee
The equations of motion can be derived by letting:
\be
r^i_\m\longrightarrow r'^i_\m=r^i_\m+\e^i_\m
\ee
which leads to the following change in the action functional:
\be
\A\longrightarrow\A'=\A+\d\A=\A+\int\! d\l\sum_{i=1}^N \frac{d\l_i}{d\l}\ \left [
\M\frac{dr^i_\m}{d\l_i}\frac{d\e^{i\m}}{d\l_i}+\half
\frac{dr^i_\m}{d\l_i}\frac{dr^{i\m}}{d\l_i}
\sum_{j=1}^N\e^j_\n\partial^{j\n}\M\right ]
\ee
and setting:
\be
\d\A=0\ \ \text{with fixed boundaries}
\ee
After some simple calculations, one can show that the equations of motion are:
\be
\frac{d}{d\l}\left (\M \frac{d\l_i}{d\l}u^i_\m\right )=\half \left (\sum_{j=1}^N\frac{d\l_j}{d\l}
u^j_\n u^{j\n}\right ) \partial^i_\m\M
\label{xx}
\ee
where:
\be
u_\m ^i=\frac{dr^i_\m}{d\l_i}
\ee
In order to investigate the symmetries of the action functional of many particle systems, we 
consider the following transformation:
\be
\l\longrightarrow\l+\d
\ee
For such $\l$--translations the condition for the action functional to be invariant 
is that:
\be
\H=\half\M\sum_{i=1}^N\frac{d\l_i}{d\l}u^i_\m u^{i\m}=E
\label{chom}
\ee
so that:
\be
\sum_{i=1}^N\frac{d\l_i}{d\l}\frac{d\t_i^2}{d\l_i^2}=\frac{2E}{\M c^2}
\ee
which on using (\ref{xx}) we have the following as the equations of motion:
\be
\M\frac{du^i_\m}{d\l_i}=\sum_{j=1}^N\left ( \half N\left (\sum_{k=1}^N u_\a^ku^{k\a}\right )
\d^{ij}\eta_{\m\n}-u_\m^i u_\n^j\right )\partial^{j\n}\M
\ee
or:
\be
\M\frac{dv^i_\m}{d\t_i}=\sum_{j=1}^N\left ( \half N^2c^2 
\d^{ij}\eta_{\m\n}-v_\m^i v_\n^j\left ( 1-\half\d^{ij}\right )\right )\partial^{j\n}\M
\ee
where:
\be
v_\m^i=\frac{dr_\m^i}{d\t_i}
\ee
In the non--relativistic limit in which $\t_i$ and $t_i$ are equal we have:
\be
\m=0\Longrightarrow \frac{d\l_i}{d\l}=\frac{1}{N}
\ee
and:
\be
\m\frac{d^2\vec{r}_i}{dt_i^2}=-\vec{\nabla}_i\left ( -\frac{\m c^2}{2}\ln (\frac{\M}{\m})\right )
\ee
or:
\be
\m\left ( \left (\frac{dt}{dt_i}\right )^2\frac{d^2\vec{r}_i}{dt^2}
+\frac{d^2t}{dt_i^2}\frac{d\vec{r}_i}{dt}\right )=
-\vec{\nabla}_i\left ( -\frac{\m c^2}{2}\ln (\frac{\M}{\m})\right )
\ee
where $t$ is the common Newtonian time. In order to have the correct non--relativistic limit
one should set:
\be
\frac{dt}{dt_i}=\text{constant}=\sqrt{\frac{m_i}{\m}}
\ee
and:
\be
\M=\m\exp\left\{\frac{\hbar^2}{\m c^2}\frac{1}{|\S|}\sum_{j=1}^N
\frac{\Box_j|\S|}{m_j}\right\}
\ee

In summary we have:
\begin{itemize}
\item For a single particle system:
\be
\M=m\exp\left\{\frac{\hbar^2}{m^2c^2}\frac{\Box|\S|}{|\S|}\right\}
\ee
\be
\M\frac{dv_\m}{d\t}=\half\left ( c^2\eta_{\m\n}-v_\m v_\n\right )\partial^\n\M
\label{REM1}
\ee
\item For many particle systems:
\be
\M=\m\exp\left\{\frac{\hbar^2}{\m c^2}\frac{1}{|\S|}\sum_{j=1}^N
\frac{\Box_j|\S|}{m_j}\right\}
\ee
\be
\M\frac{dv^i_\m}{d\t_i}=\sum_{j=1}^N\left ( \half N^2c^2 
\d^{ij}\eta_{\m\n}-v_\m^i v_\n^j\left ( 1-\half\d^{ij}\right )\right )\partial^{j\n}\M
\label{REM2}
\ee
\end{itemize}
\section{Extension to the curved space--time}
Extension of the above formulation to the case of a particle or a system of particles moving in
a curved space--time is straightforward. For a single particle, the relation (\ref{REM1}) can be generalized 
to:
\be
\M\frac{dv^\m}{d\t}+\M\G^\m_{\n\k}v^\n v^\k=\half\left ( c^2g^{\m\n}-v^\m v^\n\right )\nabla_\n\M
\ee
This is the geodesic equation of motion corrected by some quantum force. Recently\cite{Geo,Con,Stq,Geometry} it is shown that 
this equation of motion can be transformed back to the standard geodesic equation of motion 
via a specific conformal transformation. Letting the conformal transformation:
\be
\widetilde{g}_{\m\n}=\frac{\M}{m}g_{\m\n}
\ee
one gets the standard geodesic equation without  any quantum force,
in terms of the $\widetilde{g}_{\m\n}$ metric. So quantum effects are in fact geometrical effects. They determine
the conformal degree of freedom of the space--time metric.\cite{Geo,Con,Stq,Geometry}

In the case of many--particle systems,  we have the following extension of (\ref{REM2}):
\be
\M\frac{dv^{i\m}}{d\t_i}+\M\G^\m_{\n\k}v^{i\n} v^{i\k}=\sum_{j=1}^N\left (\half N^2c^2g^{\m\n}\d^{ij}
-v^{i\m}v^{j\n}\left (1-\half\d^{ij}\right )\right )\nabla^j_\n\M
\ee
It must be noted here that it is not straightforward to bring this equation into the standard 
geodesic form via a conformal transformation.
\section{Conclusion}
It is a fatal charachter of the standard relativistic de-Broglie--Bohm theory that it leads to some
mass function for the particle whose square is not positive--definite. This is in fact a result of obtaining 
the equations of the theory by decomposing the wavefunction into its phase and norm in the Klein--Gordon 
equation. But there is another view about de-Broglie--Bohm theory. One can imagine that there is an 
objectively real field in accompany to the particle satisfying an appropriate wave equation (the
Schr\"odinger equation in the non--relativistic domain and the Klein--Gordon equation for the 
relativistic case). This field exerts the quantum force on the particle.  These two pictures are
completely equivalent in the non--relativistic domain.

In the relativistic domain the first picture, i.e. decomposing the wavefunction, leads to the standard relativistic
theory which has the problem of non--positive--definitness as it is asserted above.
In the second picture, one must choose the Klein--Gordon equation as the field equation and then search for
an appropriate quantum force derivable from the wavefunction (or equivalantly, an appropriate equation of motion). 
Since in the non--relativistic
case the quantum force is known exactly, and since our relativistic equations of motion should
have the correct non--relativistic limit, our criteria for the correct equation of motion is that it be
a covariant equation which
has the correct non--relativistic limit. 

We have shown how one can write a covariant equation of motion for both a single
particle and a many particle system with the correct non--relativistic limit. 
Also the case of motion in the curved space--time is investigated.
At this end two points must be noted. First that the extension of the problem of writting down appropriate 
relativistic equations for particles with nonzero spin is also possible. This would be appeared in a forthcoming work.\cite{hehe} Second it must noted that the present theory is a covariant one with appropriate non--relativistic limit. This does not mean that the only theory satisfying these conditions is this one. There may be some other theory also covariant with appropriate limit. The experiment or some theoretical principles should decide which one is correct. But any theory of this kind should differ from the standard relativistic theory slightly. This is true with ours. Note that the higher terms of our mass function are small except for a case where the wavefunction changes with a very high frequency.

{\bf Acknowledgment}: The authors wish to thanks the anonymous referee for his or her fruitful comments.


\begin{thebibliography}{99}
\bibitem{Bohm1}
D. Bohm, Phys. Rev., {\bf 85}, 166, 1952.
\bibitem{Bohm2}
D. Bohm, Phys. Rev., {\bf 85}, 180, 1952.
\bibitem{Bohm3}
D. Bohm, and B.J. Hiley, {\it The undivided universe\/}, Routledge, 1993.
\bibitem{Holland}
P.R. Holland, {\it The quantum theory of motion\/}, Cambridge University Press, 1993.
\bibitem{de1}
L. de Broglie, {\it Les principes de la nouvelle mecanique ondulatoire\/}, J.
de Physique, Vol. 7, Series VI, 321-337, 1926.
\bibitem{de2}
 L. de Broglie, {\it La mecanique ondulatoire et la structure atomique de la
matiere et du rayonnement\/}, J. de Physique, Vol. 8, Series VI, 225-241,
1926.
\bibitem{de3}
L. de Broglie, {\it Non linear wave mechanics\/}, Elsevier, Amsterdam, 1960.
\bibitem{Thesis}
A. Shojai, PhD thesis, Sharif University of Technology, Tehran, 1997.
\bibitem{Geo}
F. Shojai, and M. Golshani, Int. J. Mod. Phys. A.,  Vol. 13, No. 4, 677, 1998.
\bibitem{Con}
F. Shojai, A. Shojai, and M. Golshani,  Mod. Phys. Lett. A.,  Vol. 13, No. 34, 2725, 1998.
\bibitem{Stq}
F. Shojai, A. Shojai, and M. Golshani,  Mod. Phys. Lett. A.,  Vol. 13, No. 36, 2915, 1998.
\bibitem{Geometry}
A. Shojai, {\it Quantum, Gravity, and Geometry\/}, {\bf IPM/P--99/046}, submitted for publication.
\bibitem{hehe}
A. Shojai, and F. Shojai, in preparation.
\end{thebibliography}
\end{document}